\documentclass[11pt]{article}

\usepackage[preprint]{acl}

\usepackage{times}
\usepackage{latexsym}

\usepackage{inconsolata}

\usepackage{microtype}
\usepackage{graphicx}
\usepackage{subcaption}
\usepackage{booktabs} 
\usepackage{hyperref}
\usepackage{natbib}
\usepackage{adjustbox}
\usepackage{enumitem}
\usepackage[utf8]{inputenc} 
\usepackage[T1]{fontenc}    
\usepackage{hyperref}       
\usepackage{url}            
\usepackage{booktabs}       
\usepackage{amsfonts}       
\usepackage{microtype}      
\usepackage{xcolor}         
\usepackage{graphicx} 
\usepackage{subcaption}
\usepackage{wrapfig}
\usepackage{multirow}
\usepackage{color}
\usepackage{colortbl}
\usepackage{algorithmic}
\usepackage{algorithm}

\renewcommand{\algorithmicrequire}{ \textbf{Input:}} 
\renewcommand{\algorithmicensure}{ \textbf{Output:}}
\usepackage{bm}
\newcommand{\ie}{\textit{i.e., }}
\newcommand{\eg}{\textit{e.g., }}

\definecolor{mygray}{rgb}{0.9, 0.9, 0.9}

\usepackage{amsmath}
\usepackage{amssymb}
\usepackage{mathtools}
\usepackage{amsthm}
\usepackage{tcolorbox}

\usepackage[capitalize,noabbrev]{cleveref}

\theoremstyle{plain}

\theoremstyle{definition}

\theoremstyle{remark}

\title{Bi-Mem: Bidirectional Construction of Hierarchical Memory for Personalized LLMs via Inductive-Reflective Agents}

\author{
Wenyu~Mao\textsuperscript{1}\thanks{Equal Contribution},\,
Haosong~Tan\textsuperscript{1}$^*$\,
Shuchang~Liu\textsuperscript{2}\,
Haoyang~Liu\textsuperscript{1},\, \\
\textbf{Yifan~Xu}\textsuperscript{\textbf{1}},\,
\textbf{Huaxiang~Ji}\textsuperscript{\textbf{1}},\,
    \textbf{Xiang~Wang}\textsuperscript{\textbf{1}}\thanks{Corresponding author: xiangwang@ustc.edu.cn.}\,\,,\,\\
\textsuperscript{1}
University of Science and Technology of China\\
\textsuperscript{2}Kuaishou Technology \\
}

\begin{document}
\maketitle

\begin{abstract}
Constructing memory from users’ long-term conversations overcomes LLMs’ contextual limitations and enables personalized interactions. Recent studies focus on hierarchical memory to model users’ multi-granular behavioral patterns via clustering and aggregating historical conversations. However, conversational noise and memory hallucinations can be amplified during clustering, causing locally aggregated memories to misalign with the user’s global persona. To mitigate this issue, we propose Bi-Mem, an agentic framework ensuring hierarchical memory fidelity through bidirectional construction.
Specifically, we deploy an inductive agent to form the hierarchical memory: it extracts factual information from raw conversations to form fact-level memory, aggregates them into thematic scenes (\ie local scene-level memory) using graph clustering, and infers users’ profiles as global persona-level memory. Simultaneously, a reflective agent is designed to calibrate local scene-level memories using global constraints derived from the persona-level memory, thereby enforcing global-local alignment. For coherent memory recall, we propose an associative retrieval mechanism: beyond initial hierarchical search, a spreading activation process allows facts to evoke contextual scenes, while scene-level matches retrieve salient supporting factual information. Empirical evaluations demonstrate that Bi-Mem achieves significant improvements in question answering performance on long-term personalized conversational tasks.
\end{abstract}

\section{Introduction}
Personalization of Large Language Models (LLMs) \citep{survey_personalized_llm_1,survey_persoanlzied_llm_2} aims to address user-specific requirements, such as recalling shared experiences and generating preference-aligned suggestions based on historical conversations. To facilitate personalized interactions in long-term conversational tasks, constructing memory \citep{Mem0, li2025cam, xu2025amem} has emerged as a critical paradigm due to the limitation of LLMs' context windows. Early work focused on extracting key entities or dialogue summaries from historical conversations as factual memories \citep{memgpt,Mem0}. However, such fragmented factual memories struggle to capture inter-fact relationships or users’ high-level behavioral patterns \citep{MemTree, pan2025secom}. Consequently, recent research has explored clustering and aggregating these isolated facts with different granularities to form a hierarchical memory structure (\eg multi-level graph \citep{li2025cam,HippoRAG} or tree structure \citep{MemTree}). By facilitating coarse-to-fine retrieval \citep{FunnelRAG}, the hierarchical systems provide more comprehensive contextual knowledge, enabling more coherent and personalized interactions.

\begin{figure}[t]
  \includegraphics[width=\columnwidth]{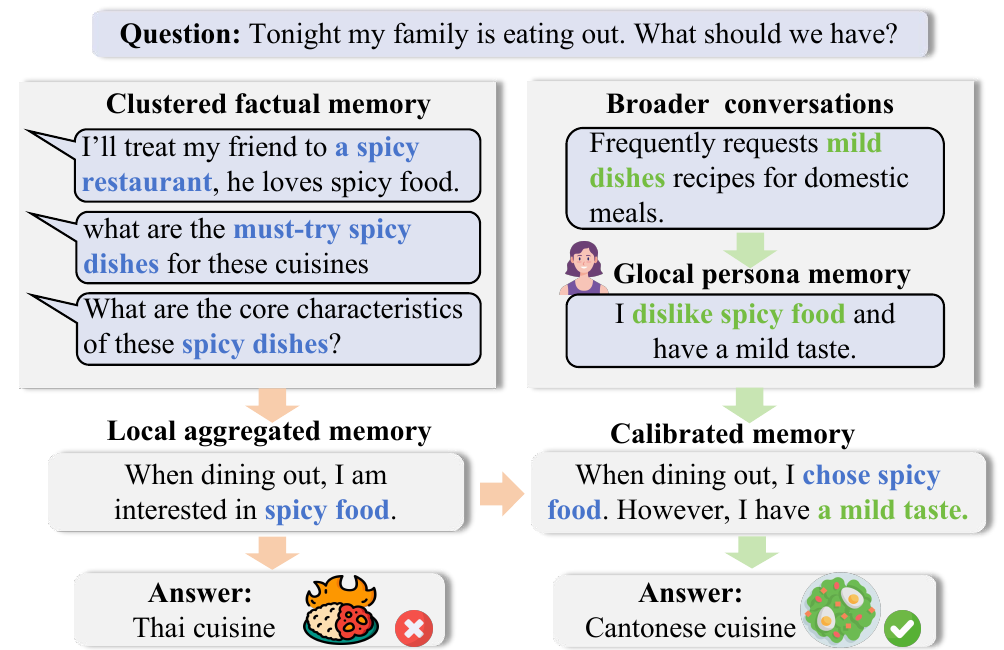}
  \caption{Illustration of local aggregated memory misaligning with the user’s global persona in naive hierarchical memory systems, leading LLMs to generate persona-violating answers.}
  \label{fig:teaser}
\end{figure}

Despite the success of hierarchical memory, such clustering and aggregation face two critical limitations: noise amplification and cumulative hallucinations \cite{llm_Hallucination, zhang2025hallucination}. First, conversational noise --- such as irrelevant chitchat and incidental interaction --- is often amplified within local clusters, causing aggregated local memory to skew toward trivialities. Second, fact-level hallucinations \citep{HaluMem} and extraction errors propagate and accumulate during hierarchical consolidation. Together, these ``cascading errors'' cause locally aggregated memories to misalign with the user’s global persona \citep{local-global}. As illustrated in Figure \ref{fig:teaser}, consider a user who frequently requests mild recipes for home-cooked meals but seeks spicy dishes for a one-off social dinner with spicy-loving friends. Naive clustering tends to amplify this transient social context, leading to the aggregation of these contexts into a local behavioral rule: ``prefers spicy cuisine when dining out.'' In contrast, the global persona derived from broader conversations is ``dislikes spicy food and has a mild taste.'' Consequently, the aggregated local memory diverges from the user’s global persona, leading LLMs to generate persona-violating suggestions for family dining-out scenarios.


To tackle these problems, we propose Bi-Mem, an agentic framework ensuring hierarchical memory fidelity via bidirectional construction, rather than unidirectional aggregation \citep{li2025cam}. Specifically, we employ an inductive agent to manage bottom-up hierarchical structure formulation: extracting atomic fact-level memory from raw conversations, aggregating them into thematic scene-level memory via graph clustering \citep{graph}, and further distilling them into a five-dimensional persona-level memory \citep{memorybank} as a global constraint. Meanwhile, the reflective agent is designed to perform top-down calibration, imposing global constraints on local scene memories to enforce global-local alignment. As shown in Figure \ref{fig:teaser}, calibrating the local scene memory by injecting the global constraint of ``I have a mild taste'' resolves the global-local contradictions, thereby enabling LLMs to generate preference-aligned suggestions.
To connect memories of different granularities, we introduce an associative retrieval mechanism. Beyond the initial hierarchical search, it adopts a spreading activation process \citep{associative}: retrieved facts evoke contextual thematic scenes, while scene-level matches retrieve salient supporting facts. Empirical results demonstrate that Bi-Mem significantly improves question answering performance in personalized long-term conversational tasks.

The main contributions of this work are summarized as follows:\begin{itemize}[leftmargin=*, itemsep=0pt, parsep=0pt, topsep=0pt, partopsep=0pt]
\item We identify global persona-local aggregated memory misalignment as a general challenge for hierarchical memory, which is induced by noise and hallucination.
\item We design Bi-Mem, an agentic framework that ensures hierarchical memory fidelity through bidirectional construction (\ie an inductive and a reflective process).
\item We propose an associative retrieval mechanism to facilitate coherent memory recall, which enhances the integration of granular factual information and contextual scenes.
\item Extensive empirical evaluations on complex personalized long-term conversational tasks demonstrate the effectiveness of Bi-Mem.
\end{itemize}
\section{Related Work}
\paragraph{Personalized LLMs.}
Personalized LLMs aim to tailor model responses to align with specific user preferences and personalities based on user-LLM interaction histories \citep{survey_personalized_llm_1,survey_persoanlzied_llm_2}. Instead of just delivering general knowledge \citep{general_LLM}, they meet user-specific needs, like recalling shared experiences or generating preference-aligned suggestions—boosting engagement and satisfaction. To handle long-term interaction histories, existing approaches either fine-tune user-specific modules \citep{finetune-persoanlized-llm,finetune-personalized-llm-2} on conversational data or adopt memory architectures \citep{local-global,memorybank} to store and manage user-specific conversations without modifying model parameters. Considering the computational resource of fine-tuning and LLMs’ context window limitation, we focus on building memory for users’ long-term conversational history to enable such personalization.

\paragraph{Memory for Personalized LLMs.}
Personalized LLMs' memory \citep{zhang2024survey,hu2025memory} is a flexible component that stores and leverages users' historical conversation information to supplement the LLMs’ context window limitations. It mainly consists of two parts: memory construction \citep{xu2025amem}(extracting key information from user-LLM interactions and managing it into structured forms) and memory retrieval \citep{retrieval}(retrieving relevant memory through semantic similarity to support LLMs' response generation). Mainstream implementations adopt token-based storage. For example, Mem0 \citep{Mem0} constructs memory from conversational information by supporting ADD/UPDATE/DELETE/NOOP operations and retrieves relevant memory through vector similarity.
MemoryBank \citep{memorybank} empowers personalized LLMs by integrating human-like memory storage (dialogues, event summaries, user portraits) and dense retrieval.

\paragraph{Hierarchical Memory.}
Token-based memory for personalized LLMs has two core structures: Flat Memory \citep{memgpt} stores information as independent units (\eg raw interaction and summarized sessions) without explicit connections. Hierarchical Memory \citep{hu2025memory} uses multi-level architectures with different granularities and associations, supporting complex reasoning via cross-layer links. For example, CAM \citep{li2025cam} constructs hierarchical memory by adopting an incremental overlapping clustering algorithm and retrieves by a Prune-and-Grow strategy. HippoRAG \citep{HippoRAG} employs a hierarchical schemaless knowledge graph (KG) for memory construction and leverages Personalized PageRank for cross-memory reasoning.

\section{Preliminary}

\subsection{Task Formulation}
Let $\mathcal{C} = \{c_1, ..., c_N\}$ denote the long-term conversational history between a user and an LLM agent, where each $c_i = (q_i, r_i)$ denotes the $i$-th interaction (user request $q_i$, LLM response $r_i$). The task of the memory agent is composed of a memory construction and retrieval task \citep{pan2025secom}. 

\begin{itemize}[leftmargin=*, itemsep=0pt, parsep=0pt, topsep=0pt, partopsep=0pt]
\item \textbf{Memory construction:} Construct a structured memory bank $\mathcal{M}$ by encoding conversational history $\mathcal{C}$ via a memory construction function:
\begin{gather}
\mathcal{M} = f_{\text{cons}}(\mathcal{C})
\end{gather}
where $f_{\text{cons}}$ constructs $\mathcal{C}$ into discrete memory units $\boldsymbol{m} \in \mathcal{M}$.  
\item \textbf{Memory Retrieval:} Given a target user query $q^*$ and a memory bank $\mathcal{M}$, extract $K$ memory units relevant to $q$ via a retrieval function $f_R$:
\begin{gather}
\{ \boldsymbol{m}_k\in \mathcal{M} \}_{k=1}^K \leftarrow f_R(q^*, \mathcal{M}, K)
\end{gather}

\item \textbf{Response Generation:} The goal of LLM personalization \citep{local-global} is to generate an optimal personalized response $r^*$ for the current query $q^*$ based on retrieved $N$ memory units:
\begin{gather}
r^* = f_{\text{LLM}}\left(q^*, \{ \boldsymbol{m}_k \}_{k=1}^K \right)
\end{gather}
\end{itemize}
where $r^*$ aligns with the user’s persona or accurately recalls relevant contextual details.

\section{Method}
\begin{figure*}[t]
  \includegraphics[width=0.95\textwidth]{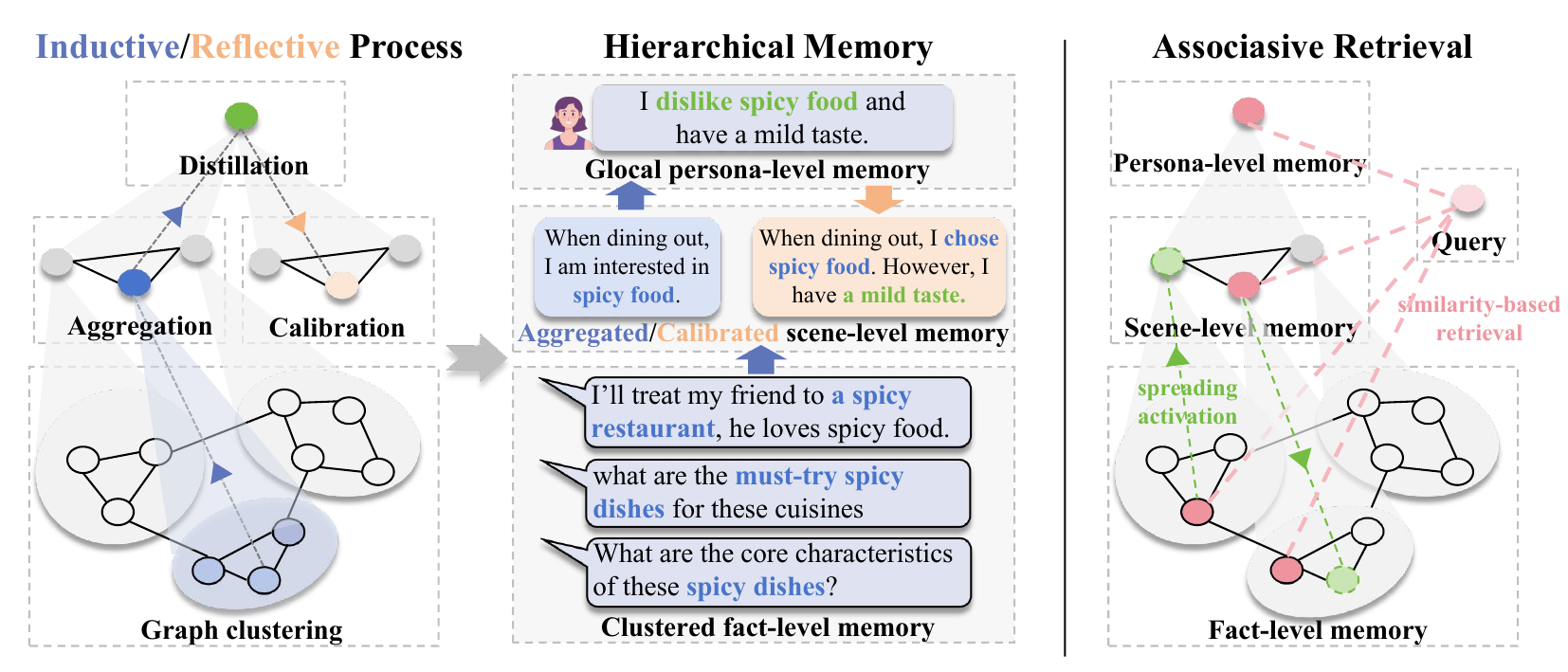}
  \centering
  \caption{The framework of our proposed Bi-Mem, including the bidirectional construction (inductive process and reflective process) of hierarchical memory and the associative retrieval.}
  \label{fig:method}
\end{figure*}
In this section, we elaborate on our proposed Bi-Mem, an agentic framework with bidirectional construction for hierarchical memory and associative retrieval, as shown in Figure \ref{fig:method}. We first define the three levels of our designed hierarchical memory in \ref{definition}. Then, we present bidirectional memory construction in Section \ref{Memory_Construction}, including the inductive-reflective process, followed by the details of the associative retrieval mechanism in Section \ref{Memory_Retrieval}. The detailed algorithm is presented in Appendix \ref{app: algo}.

\subsection{Definition for Hierarchical Memory}
\label{definition}
The hierarchical memory $\mathcal{M} = (\mathcal{F}, \mathcal{S}, \mathcal{P})$ in Bi-Mem is structured into three granularities:
\begin{itemize}[leftmargin=*, itemsep=0pt, parsep=0pt, topsep=0pt, partopsep=0pt]
\item \textbf{Fact-level memory} ($\mathcal{F}$) is the atomic unit of memory, where each conversational interaction maps to a single fact unit. Formally, $\mathcal{F} = \{f_1, f_2, ..., f_{N}\}$, where each unit $f_i$ is defined as the tuple $(i,\text{con}_i, t_i, e_i)$: $i$ is the index of fact, $\text{con}_i$ is the summarized content of the interaction, $t_i$ is the timestamp, and $e_i$ is the connection edges linking to related fact or interaction indices.

\item \textbf{Scene-level memory} ($\mathcal{S}$) clusters related facts to form a thematic scene (\eg ``dining-out with spicy-loving friends''). Formally, $\mathcal{S} = \{s_1, s_2, ..., s_J\}$, where each scene $s_j$ is the aggregation of a subset of fact-level memory $\bar{\mathcal{F} }_j \subseteq \mathcal{F}$: $s_j = \frac{1}{|\bar{\mathcal{F} }_j |} \sum_{f \in \bar{\mathcal{F} }_j } f$.
  
\item \textbf{Persona-level memory} ($\mathcal{P}$) represents the user’s core profile, distilled from all aggregated scenes $\mathcal{S}$. It serves as a global constraint to ensure memory consistency. Formally, $\mathcal{P} = [p_1, p_2, p_3, p_4, p_5]$ is a five-dimensional persona, where $p_1$ (basic-info) covers name, age, occupation, and location, $p_2$ (interests) covers hobbies, likes, and dislikes, $p_3$ (personality) covers personality traits, emotional patterns, $p_4$ (values) covers core values, beliefs, and motivations, and $p_5$ (relationships) covers key social relationships. 
\end{itemize}

\subsection{Bidirectional Memory Construction}
\label{Memory_Construction}
Bidirectional construction integrates a bottom-up inductive process (for memory formulation) and a top-down reflective process (for top-down calibration), realizing the global-local alignment.

\subsubsection{Inductive Process (Bottom-Up)}
The inductive agent constructs the hierarchical memory from raw conversations through three sequential steps:

\paragraph{Fact-level Memory Extraction:} To construct the fact unit $f_i = (i, \text{con}_i, t_i, e_i)$ for each interaction $c_i$ in the conversation, we first apply a summarization operator $\mathcal{E}$ via an LLM to extract the summarized interaction content $\text{con}_i$, defined as:
\begin{gather}\text{con}_i = \arg\max_{\text{con} \in \text{Con}} P_\mathcal{E}(\text{con} | c_i), 
\end{gather}
where $i \in[1,2,\ldots,N]$, and $\text{Con}$ is the candidate set of generated summaries by LLMs. The index $i$ and timestamp $t_i$ are directly assigned from the interaction metadata. The connection edges $e_i$ are identified by computing the semantic similarity between the current content $\text{con}_i$ and $\text{con}_l$ from other fact units. Formally, $l \in e_i$ if:
\begin{gather}
\text{sim}(\phi(\text{con}_i), \phi(\text{con}_l)) > \tau, 
\label{equ: edge}
\end{gather}
where $\text{sim}(\cdot)$ is the cosine similarity of embeddings obtained by a pretrained embedding model $\phi$, and $\tau$ is a predefined threshold.

\paragraph{Scene-Level Memory Aggregation:} To provide more contextual knowledge and capture users' behavioral patterns, we transform fragmented atomic facts into high-level thematic scenes. We model the factual memory as an undirected graph $G = (\mathcal{F}, E)$, where an edge $(f_i, f_l) \in E$ is established if $l \in e_i$ or $i \in e_l$. We then perform graph clustering by applying the Label Propagation Algorithm (LPA) \citep{LPA} to $G$. This process utilizes local structural consensus \cite{lancichinetti2012consensus} to partition the fact set $\mathcal{F}$ into $J$ thematic clusters $\{\bar{\mathcal{F}}_1, \bar{\mathcal{F}}_2, \dots, \bar{\mathcal{F}}_J\}$. Formally, for each fact cluster $\bar{\mathcal{F}}_j \subseteq \mathcal{F}$, we apply an aggregation operator $\mathcal{A}$ with LLMs to derive the corresponding scene $s_j$ by maximizing the generation probability:

\begin{gather} s_j = \arg\max_{s \in \mathcal{S}_{can}} P_{\mathcal{A}}(s \mid {f_k \in \bar{\mathcal{F}}_j }), 
\end{gather}
where $\mathcal{S}_{can}$ is the set of candidate generated scenes. Thus, we can construct thematic scene-level memory $\mathcal{S} = \{s_1, s_2, \dots, s_J\}$.

\paragraph{Persona-Level Memory Distillation:} To obtain the global constraint of memory, we distill the five-dimensional persona $\mathcal{P}$ from the entire scene-level memories, as detailed in \ref{definition}. Specifically, we define a distillation operator $\mathcal{D}$ that leverages an LLM to project the set $\mathcal{S}$ onto each persona dimension:
\begin{gather}
p_d = \arg\max_{p \in \mathcal{P}_{cand}} P_{\mathcal{D}}(p \mid \mathcal{S}, \text{per}_d),
\end{gather}
where $d \in {1, \dots, 5}$, $\text{per}_d$ is a dimension-specific instruction that guides the LLM to focus on relevant scenes within $\mathcal{S}$. 

\subsubsection{Reflective Process (Top-Down)}
To enforce global-local alignment and mitigate cascading errors from the inductive process, we implement a reflective agent that performs top-down calibration, ensuring that each local scene $s_j \in \mathcal{S}$ is grounded in the stable persona constraints $\mathcal{P}$.

Specifically, the reflective agent first assesses whether the local scene $s_j$ contradicts or fails to reflect the related global persona within $\mathcal{P}$. It then applies a Calibration Operator $\mathcal{R}$ to generate a compensatory condition $\Delta s_j$ that explicitly bridges the local scene $s_j$ with the global persona:
\begin{gather}
\Delta s_j = \arg\max_{\delta \in \Delta_{cand}} P_{\mathcal{R}}(\delta \mid s_j, \mathcal{P}),
\end{gather}
where $\Delta_{cand}$ denotes the space of candidate compensatory conditions. The calibrated scene is formally defined as: $s_j' = s_j \oplus \Delta s_j$, where $\Delta s_j$ is empty if the local scene is already consistent with the global persona. This reflective process injects global constraints on local scenes, resulting in a self-consistent hierarchical memory with scenes $\mathcal{S}' = \{s_1', s_2', \dots, s_J'\}$. Since fact-level memory consists of detailed conversation facts, we refrain from calibrating it to align with reality.

\subsection{Associative Memory Retrieval}
\label{Memory_Retrieval}
To enable coherent recall across memory granularities, we propose an associative retrieval mechanism. This mechanism leverages Spreading Activation after initial search in hierarchical memory $\mathcal{M} = (\mathcal{F}, \mathcal{S}', \mathcal{P})$, balancing granular detail with high-level context.
\paragraph{Initial Hierarchical Search:} Given a query $q^*$, we first project both the query and all memories into a latent space using a pre-trained embedding model $\phi(\cdot)$. Initial relevance scores $a_0$ are computed as the cosine similarity between the query and each memory unit:
\begin{gather}
\label{eq: initial_search}
a_0(x) = \text{sim}(\phi(q^*),\phi(x)), 
\end{gather}
where $x \in \mathcal{F} \cup \mathcal{S}' \cup \mathcal{P}$ is the memory units in hierarchical memory, $\text{sim}(\cdot)$ is the cosine similairty. The top-$k$ $x$ with the highest $a_0$ scores are selected as the retrieved memory set $M_{ret}$.

\paragraph{Associative Spreading Activation:}
To capture structural dependencies between granular facts and thematic scenes, we implement a bidirectional associative mechanism: the retrieved fact invokes its parent scene, while the retrieved scene triggers the recall of its most representative facts. Specifically, starting from the initial retrieved memory set $\mathcal{M}_{ret}$, the retrieval set is expanded through two associative paths. First, for each fact $f \in \mathcal{M}_{ret}$, its unique parent scene $s$ is automatically associated and added to the expanded set. Second, for each scene $s_j \in \mathcal{M}_{ret}$, the mechanism spreads activation to its corresponding fact cluster $\bar{\mathcal{F}}_j$. We select the top-$m$ facts with the highest relevance score $a_1$ to the scene:
\begin{gather}
\label{eq: associative}
a_1(x) =\text{sim}(\phi(s_j),\phi(x)), \quad \forall x \in \bar{\mathcal{F}}_j.
\end{gather}
And for each persona $p \in \mathcal{M}_{ret}$, we treat it as a global anchor and do not associate it with other memories.  
The final retrieved memory set $\mathcal{M}_{ret}$ is updated as the union of the initial retrieved ones and those associated ones. 

\paragraph{Response Generation:}
The final retrieved memory set $\mathcal{M}_{ret}$, which encompasses multi-granular evidence from atomic facts to global persona traits, is integrated into a unified prompt context. Formally, given the current query $q^*$ and the retrieved information in $\mathcal{M}_{ret}$, the LLMs generates a personalized response $r^*$ by maximizing the conditional probability:
\begin{gather}
r^* = \arg\max P_{\text{LLM}}(r \mid q^*, \mathcal{M}_{ret}).
\end{gather}

\section{Experiments}

\begin{table*}[t!]
    \resizebox{\textwidth}{!}{%
    \begin{tabular}{cc|cccccccccc}
    \toprule
    \multirow{2}{*}{\textbf{Model}} & \multirow{2}{*}{\textbf{Method}} & \multicolumn{8}{c}{\textbf{LoCoMo}} \\
    \cmidrule(l){3-12} 
     & & \multicolumn{2}{c}{\textbf{Single Hop}} & \multicolumn{2}{c}{\textbf{Multi Hop}} & \multicolumn{2}{c}{\textbf{Temporal}} & \multicolumn{2}{c}{\textbf{Open Domain}} & \multicolumn{2}{c}{\textbf{Average}} \\
     & &  $F_1 \uparrow$ & $B_1 \uparrow$   & $F_1 \uparrow$ & $B_1 \uparrow$  &  $F_1 \uparrow$ & $B_1 \uparrow$  & $F_1 \uparrow$ & $B_1 \uparrow$  & $F_1 \uparrow$ & $B_1 \uparrow$\\
    \midrule
    \multirow{7}{*}{\rotatebox{90}{\textbf{GPT-4o-mini}}} 
     & \textsc{LongContext} & 32.69  & 20.85 & 22.87 & 16.84 & 12.47 & 8.86 & 7.87 & 5.89 & 25.08 & 16.65 \\
    \cmidrule{2-12}
     & \textsc{Rag} & \underline{52.45} & \underline{47.94} & 27.50 & 20.13 & 46.07 & 40.35 & 23.23 & 17.94 & 44.67 & \underline{39.33} \\
     & \textsc{SeCom} &  16.67 & 13.96 & 15.06 & 11.93 & 14.58 &12.20 & 17.29 & 14.76 & 15.97 & 13.25  \\
     & \textsc{A-Mem} & 44.65 & 37.06 & 27.02 & 20.09 & 45.85 & 36.67 & 12.14 & 12.00 & 39.61 & 32.27 \\
     & \textsc{Mem0} & 47.65 & 38.72 & \underline{38.72} & \underline{27.13} & \underline{48.93} & \underline{40.51} & \underline{28.64} & \underline{21.58} & \underline{45.08} & 35.88 \\
     & \textsc{LightMem} & 41.79 & 37.83 & 29.78 & 24.80 & 43.71 & 39.72 & 16.89 & 13.92 & 38.41 & 34.32 \\
     & \textsc{CAM} &  49.94  & 46.68  & 36.63 & 25.37 & 18.85 & 12.90 & 22.71 & 18.93 & 39.25 & 33.92 \\
    \cmidrule{2-12}
     & \textbf{Bi-Mem} & \textbf{53.68} & \textbf{47.99} & \textbf{39.17} & \textbf{31.81} & \textbf{54.44} & \textbf{41.56} & \textbf{30.99} & \textbf{26.90} & \textbf{49.74} & \textbf{42.33} \\
    \midrule
    \multirow{7}{*}{\rotatebox{90}{\textbf{Qwen2.5-14b-Instruct}}} 
     & \textsc{LongContext} & 21.00  & 13.17 & 18.09  & 13.21 & 10.54  & 5.82 & 7.31 & 4.71 & 17.41 & 11.10 \\
    \cmidrule{2-12}
     & \textsc{Rag} & \underline{47.87} & \underline{42.79} & 26.38 & 19.54 & 30.78 & 25.97 & 14.16 & 10.52 & \underline{38.20} & \underline{32.95} \\
     & \textsc{SeCom} & 22.26 & 18.71 & 19.15 & 14.26 & 19.58 & 16.30 & 15.75 & \underline{13.01}  &  20.71 & 17.02   \\  
     & \textsc{A-Mem} & 33.75 & 30.04 & 22.09 & 15.28 & 27.19 & 22.05 & 13.49 & 10.74 & 28.95 & 24.43 \\
     & \textsc{Mem0} & 42.58 & 35.15 & \underline{31.73} & \underline{24.82} & 28.96 & 26.24 & 15.03 & 11.28 & 35.99 & 29.88 \\
     & \textsc{LightMem} & 34.92 & 31.22 & 25.45 & 19.61 & \underline{32.03} & \underline{27.70} & \underline{15.81} & 11.81 & 31.37 & 27.12 \\
     & \textsc{CAM} &  32.69 & 23.15 & 24.78   & 18.98  &  5.21   & 2.71 & 15.25  & 12.45 & 24.37 & 17.42 \\
    \cmidrule{2-12}
     & \textbf{Bi-Mem} & \textbf{48.07} & \textbf{42.95} & \textbf{32.42} & \textbf{25.53} & \textbf{44.21} & \textbf{29.29} & \textbf{18.38} & \textbf{16.11} & \textbf{42.51} & \textbf{35.19} \\
    \bottomrule
    \end{tabular}%
    }
    \caption{ Overall performance of different methods on LoCoMo benchmarks. The highest score is typed in bold to indicate statistically significant improvements (p < 0.05), while the second-best score is underlined.}
    \label{tab:main_results}
\end{table*}

In this section, we conduct extensive experiments to evaluate the effectiveness of our proposed framework in long-term personalized conversational tasks by answering the following questions: 
\begin{itemize}[leftmargin=*, itemsep=0pt, parsep=0pt, topsep=0pt, partopsep=0pt]
\item RQ1: How does our framework Bi-Mem perform compared with leading baselines of memory? 
\item RQ2: What contribution does each of the Fact-Scene-Persona levels and reflective calibration make to constructing the Hierarchical Memory?
\item RQ3: To what extent does the Associative Retrieval mechanism improve the QA accuracy?
\item RQ4: How sensitive is the framework to the hyperparameters of initial retrieval size $k$?
\item RQ5: How efficient is the Bi-Mem compared to existing memory frameworks?
\end{itemize}

\subsection{Expermental Settings}
\paragraph{Datasets.} We adopt LoCoMo \citep{LOCOMO} to conduct our experiments, which is a widely-used dataset in personalized long-term conversational tasks. Consisting of 50 dialogues, each averaging 305 turns, 20 sessions, and 9,000 tokens, it offers an advantage over existing conversational datasets \citep{benchmarks_old, benchmark_msc} for constructing LLMs' memory. LoCoMo is built through a rigorous human-LLM co-creation pipeline: LLM-based agents with unique personas generate initial dialogues, and human annotators refine them to ensure long-term consistency. For further question answering based on the historical conversations (constructed memory), LoCoMo includes 7,512 question-answer pairs spanning different types: single-hop questions derived from a single session, multi-hop questions requiring cross-session information synthesis, temporal reasoning questions assessing the grasp of time-related cues, open-domain knowledge questions integrating conversational context with external facts, and adversarial questions designed to test models’ ability to recognize unanswerable queries. Note that the adversarial question category is not adopted in our work, as it does not align with the scope of memory construction and retrieval.

\paragraph{Baselines}
We evaluate the performance of Bi-Mem against multiple leading baselines thoroughly, including: LongContext (LLM backbone without memory), RAG \citep{RAG}, Mem0 \citep{Mem0}, LightMem \citep{lightmem}, A-MEM \citep{xu2025amem}, SeCom \citep{pan2025secom}, and CAM \citep{li2025cam}. The detailed introduction for these baselines is presented in the Appendix \ref{app: baseline}.

\begin{figure}[t!]
    \centering
    \includegraphics[width=1.0\linewidth]{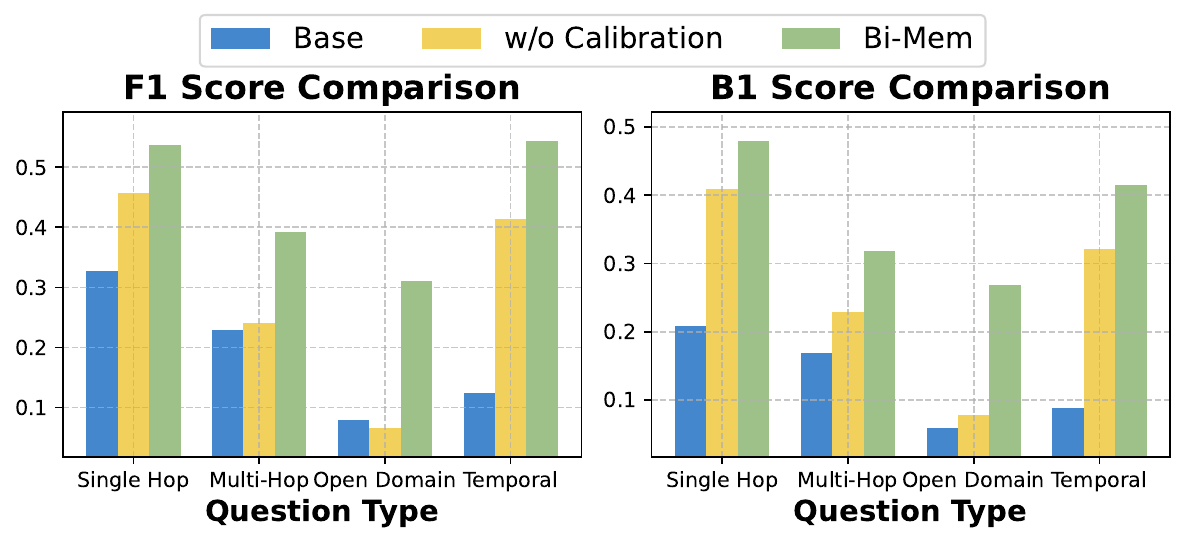}
    \caption{Ablation Study on Reflective Calibration in Hierarchical Memory Construction. ``Base'' denotes answer generation via long-context LLM backbones without memory, while ``w/o Calibration'' refers to unidirectional hierarchical memory construction without calibration in the reflective process.}
    \vspace{-2mm}
    \label{fig:abla_cali}
\end{figure}

\begin{figure}[t]
    \centering
    \includegraphics[width=1.0\linewidth]{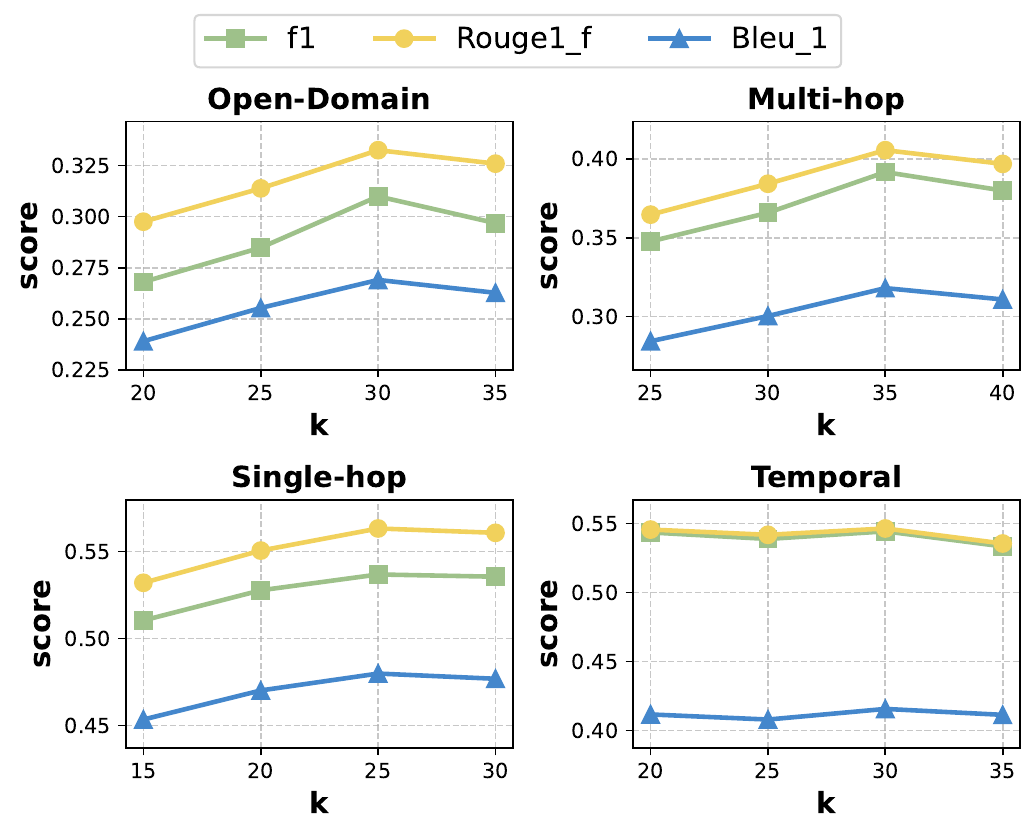}
    \vspace{-4mm}
    \caption{Sensitivity analysis of Bi-Mem to the hyperparameter $k$, which is the number of initially retrieved memory units in the hierarchical search stage.}
    \vspace{-4mm}
    \label{fig:sensitivity_k}
\end{figure}

\begin{table*}[t!]
    \resizebox{\textwidth}{!}{%
\begin{tabular}{l|cc|cc|cc|cc|cc}
    \toprule
    \multirow{2}{*}{\textbf{Method}} & \multicolumn{2}{c|}{\textbf{Single Hop}} & \multicolumn{2}{c|}{\textbf{Multi-Hop}} & \multicolumn{2}{c|}{\textbf{Open Domain}} & \multicolumn{2}{c|}{\textbf{Temporal}} & \multicolumn{2}{c}{\textbf{Average}} \\
     & $F_1 \uparrow$ & $B_1 \uparrow$ & $F_1 \uparrow$ & $B_1 \uparrow$ & $F_1 \uparrow$ & $B_1 \uparrow$ & $F_1 \uparrow$ & $B_1 \uparrow$ & $F_1 \uparrow$ & $B_1 \uparrow$ \\
    \midrule
    Fact only & 49.99 & 44.66 & 27.11 & 20.46 & 9.43 & 7.80 & 51.28 & 36.60 & 43.49 & 36.19 \\
    Fact+Scene & 50.12 & 44.66 & 32.01 & 23.56 & 25.75 & 22.08 & 52.35 & 37.30 & 45.72 & 37.81 \\
    Fact+Persona & 49.26 & 43.99 & 27.91 & 20.60 & 9.65 & 9.21 & 51.84 & 37.09 & 43.38 & 36.04 \\
    \textbf{Bi-Mem} & \textbf{53.68} & \textbf{47.99} & \textbf{39.17} & \textbf{31.81} & \textbf{30.99} & \textbf{26.90} & \textbf{54.44} & \textbf{41.56} & \textbf{49.74} & \textbf{42.33}\\
    \bottomrule
\end{tabular}
    }
    \caption{Ablation study to evaluate the respective contributions of the Fact-Scene-Persona levels in the hierarchical memory. "Fact only" denotes memory constructed solely at the fact level; "Fact+Scene" and "Fact+Persona" represent hierarchical memory combining fact-level with scene-level or persona-level information.}
    \label{tab:multi-granularity}
\end{table*}

\begin{table*}[t!]
    \centering
    \resizebox{\textwidth}{!}{%
    \begin{tabular}{l|cc|cc|cc|cc|cc}
        \toprule
        \multirow{2}{*}{\textbf{Method}} & \multicolumn{2}{c|}{\textbf{Single Hop}} & \multicolumn{2}{c|}{\textbf{Multi-Hop}} & \multicolumn{2}{c|}{\textbf{Open Domain}} & \multicolumn{2}{c|}{\textbf{Temporal}} & \multicolumn{2}{c}{\textbf{Average}}\\
         & $F_1 \uparrow$ & $B_1 \uparrow$ & $F_1 \uparrow$ & $B_1 \uparrow$ & $F_1 \uparrow$ & $B_1 \uparrow$ & $F_1 \uparrow$ & $B_1 \uparrow$ & $F_1 \uparrow$ & $B_1 \uparrow$ \\
    \midrule
    Top-Down & 13.87 & 11.85 & 12.92 & 9.51 & 14.47 & 13.43 & 16.63 & 11.58 & 14.32 & 11.46 \\
    Botom-Up& 42.59 & 37.50 & 29.01 & 20.95 & 21.26 & 17.03 & 47.05 & 33.26 & 39.68 & 32.27 \\
    \midrule
    Hierarchical & 51.26 & 45.76 & 32.28 & 24.58 & 20.17 & 17.46 & 51.42 & 36.16 & 45.84 & 38.06\\
    +Scene\_to\_Fact & 52.97 & 47.31 & 34.02 & 26.81 & 22.72 & 19.38 & 53.48 & 38.05 & 47.68 & 39.83 \\
    +Fact\_to\_Scene & 52.40 & 46.75 & 34.16 & 26.96 & 25.14 & 20.51 & 51.62 & 36.44 & 47.16 & 39.29\\
    \textbf{Bi-Mem} & \textbf{53.68} & \textbf{47.99}  & \textbf{39.17} & \textbf{31.81} & \textbf{30.99} & \textbf{26.90} & \textbf{54.44} & \textbf{41.56}  & \textbf{49.74} & \textbf{42.33} \\
    \bottomrule
    \end{tabular}%
    }
    \caption{Ablation study on different retrieval strategies. ``Top-down'' and ``Bottom-up'' denote retrieving relevant memories from the top or bottom level, followed by level-by-level association. ``Hierarchical'' refers to retrieving the top-k relevant memories from all three levels simultaneously. ``+ Scene\_to\_Fact'' and ``+ Fact\_to\_Scene'' extend hierarchical retrieval by spreading from scenes to their child facts or from facts to their parent scenes.}
    \label{tab:retrieval_strategies}
\end{table*}

\paragraph{Implementation Details}
For a fair comparison, we apply the same LLM backbones and embedding models for both our Bi-Mem and all baselines. Specifically, we adopt GPT-4o-mini \citep{hurst2024gpt} and Qwen2.5-14B-Instruct \citep{bai2025qwen2} (128K-token context window) to assess Bi-Mem’s effectiveness and generalizability. For embedding models, we use all-MiniLM-L6-v2 (denoted $\phi(\cdot)$, which supports both graph edge construction (for fact-level memory) and retrieval tasks. The predefined threshold $\tau$ for edge formation between fact units (Eq. \eqref{equ: edge}) is set to 0.2. Regarding the retriever implementation, a hybrid retrieval approach is adopted, which fuses a cosine similarity-based dense retriever and a BM25 retriever with the weighting factor of $0.5$. The $k$ for the initial search in Eq \eqref{eq: initial_search} is selected in $[15, 20, 30,35,40]$, while $m$ for associative spreading in Eq \eqref{eq: associative} ranges from $[1,2,3]$. The optimal hyperparameter settings are presented in the Appendix \ref{app: hyper}. To quantify the memory's effectiveness, we evaluate all methods by comparing their QA accuracy using two metrics $F_1$ (balancing answer precision and recall) and BLEU-1 (denoted $B_1$, which measures word overlap between generated and ground-truth answers).

\subsection{Main Results (RQ1)}
To answer Q1, we compare Bi-Mem with leading baselines on four types of questions from the LoCoMo benchmark. We conducted each experiment at least three times on different random seeds, with the averaged results reported in Table \ref{tab:main_results}. We can observe that Bi-Mem consistently outperforms all baselines on both GPT-4o-mini and Qwen-2.5-14B-Instruct backbones, validating the effectiveness and generalizability of our bidirectional memory construction and associative retrieval for LLMs' personalization.
Notably, RAG \citep{RAG} serves as flat factual memory in our setup, only outperforming on single-hop questions. This demonstrates that purely factual memory fails to handle complex questions requiring inter-fact relationship capture or high-level user behavioral pattern recognition, highlighting the necessity of a multi-granular memory structure.

\subsection{Ablation for Memory Construction (RQ2)}

To address Q2, we conduct experiments to verify the necessity of each granular memory in the hierarchical structure. As shown in Table \ref{tab:multi-granularity}, fact-only memory construction exhibits the worst performance across all question types except single-hop, highlighting the importance of high-level memory. Benefiting from persona-level calibration that provides global constraints, ``Fact+Scene'' achieves comparable performance. Furthermore, Bi-Mem (integrating all three levels) attains the best performance, confirming the necessity of each granularity. To further evaluate the effectiveness of bidirectional construction, we conduct an ablation study focused on reflective process calibration. As illustrated in Figure \ref{fig:abla_cali}, removing the reflective process (w/o Calibration) results in significant performance degradation across all question types. Additionally, we present a case in the Appendix \ref{app: case} to validate the significance of reflective calibration vividly.

\subsection{Ablation for Associative Retrieval (RQ3)}
\label{sec: abla_retriever}
To evaluate the effectiveness of our Associative Retrieval mechanism, we compare it with various retrieval strategies as presented in Table \ref{tab:retrieval_strategies}. The results indicate that unidirectional strategies exhibit clear limitations in bridging multi-granular memories. ``Top-down'' strategy may fail to retrieve the precise details grounded in the questions, and ``Bottom-Up'' strategy may frequently suffer from semantic fragmentation. The ``Hierarchical'' baseline, which retrieves the top-$k$ candidates from all levels simultaneously, improves performance but remains constrained by the independence of each retrieved memory unit. The significant performance gains observed with "+ Fact\_to\_Scene" and "+ Scene\_to\_Fact" demonstrate the power of spreading activation after initial hierarchical search. Ultimately, the full Bi-Mem retrieval configuration achieves the best results, proving that bidirectional association effectively bridges the gap between abstract persona and concrete conversational facts.

\subsection{Sensitivity Analysis (RQ4)}
To address RQ4, we analyze Bi-Mem’s sensitivity to the hyperparameter $k$ (number of initially retrieved memory units in the hierarchical search stage). $k$ is varied over $[15, 20, 25, 30, 35, 40]$. As shown in Figure \ref{fig:sensitivity_k}, answer accuracy across all tasks initially increases with $k$ --- indicating that a larger initial retrieval budget enables our associative retrieval to capture a more comprehensive memory set. Performance then degrades after peaking, as excessive retrieved memory introduces additional noise that negatively impacts relevant information extraction. This underscores that an optimal $k$ effectively bridges gaps between fragmented facts and thematic scenes. More sensitivity analysis on other hyperparameters (\eg $m$ for associative spreading in Eq. \ref{eq: associative}) is listed in the Appendix \ref{app: sen}.

\begin{table}[t!]
    \resizebox{\columnwidth}{!}{%
    \begin{tabular}{lccc}
    \toprule
      \textbf{Metric} & \textbf{A-mem} & \textbf{CAM} & \textbf{Bi-Mem} \\
    \midrule
     Memory Construction (s) & 3327 & 443 & 3448\\
     Retrieving-Answering (s) & 7.30 & 8.86 & 4.54  \\
     \textsc{Total} (s) & 4437 & 1790 & 4138 \\
     \midrule
     \textsc{Answer Quality} (F1) & 39.65 & 39.25 & 49.73 \\
    \bottomrule
    \end{tabular}%
    }
    \caption{Cost-Efficiency Analysis: We measured the memory construction time and average retrieving-answering time per question for a single user, with total time encompassing memory construction and answering the subsequent 152 questions. }
    \label{time_comparison}
\end{table}

\subsection{Efficiency (RQ5)}
To evaluate efficiency, we compare Bi-Mem’s memory construction and QA time costs with memory-based baselines (A-Mem \citep{xu2025amem}, CAM \citep{li2025cam}). As shown in Table \ref{time_comparison}, Bi-Mem has superior QA efficiency (4.54s/question), significantly faster than A-Mem (7.30s) and CAM (8.86s). This speedup stems from our Associative Retrieval mechanism—pre-established hierarchical links enable faster multi-granular evidence identification than flat retrieval or complex re-ranking.
Though its bidirectional Inductive-Reflective process increases construction time (3,448s) vs. CAM (443s), it matches A-Mem’s efficiency (3,327s). Notably, this trade-off is justified by large accuracy gains, confirming the value of bidirectional hierarchical memory construction.

    
    

\section{Conclusion}
In this paper, we present Bi-Mem, a framework designed to ensure hierarchical memory fidelity via bidirectional construction and an associative retrieval mechanism. Instead of constructing hierarchical memory through undirectional aggregation, Bi-Mem employs an inductive agent for bottom-up formulation and a reflective agent for top-down calibration, mitigating the misalignment between local memory and the user’s global persona. To ensure coherent recall, the associative retrieval mechanism connects memories of different granularities via spreading activation. Experimental results demonstrate that Bi-Mem significantly enhances QA accuracy in long-term personalized conversational tasks.

\clearpage
\section*{Limitations}
While Bi-Mem demonstrates significant improvements in the memory system for personalized LLMs, there are several limitations:

1. Sensitivity to Model Reasoning: The fidelity of bidirectional memory construction is inherently tied to the instruction-following and reasoning capabilities of the underlying LLM. Variations in LLMs' instructions may impact the effectiveness of the inductive and reflective processes, suggesting a need for more robust prompting strategies.

2. Static vs. Dynamic Persona Modeling: While our framework ensures consistency with a stable persona anchor, it is primarily designed for users with relatively stable personas. Future work could extend the calibration mechanism to better capture the temporal evolution of user preferences and dynamic persona shifts.

3. Potential for Optimization via RL: Our current framework focuses on the structural design of memory construction and retrieval. While effective, there is an opportunity to further refine agent-specific behaviors, such as the reflective agent’s calibration policy, through Reinforcement Learning (RL) or preference optimization to better align with diverse user interaction styles.

\section*{Ethics Statement.}
This work is designed to enhance the fidelity of hierarchical memory systems, formulating the memory construction process as a bidirectional alignment between local scenes and global personas. The generative AI is used for coding and writing assistance. We do not foresee any direct, immediate, or negative societal impacts of our research. 

\section*{Reproducibility Statement.}
All the results in this work are reproducible. We have discussed the optimal hyperparameters and the details on devices and software environments in Appendix \ref{app: hyper}.

\bibliography{reference}

\begin{thebibliography}{33}
\providecommand{\natexlab}[1]{#1}

\bibitem[{Bai et~al.(2025)Bai, Chen, Liu, Wang, Ge, Song, Dang, Wang, Wang, Tang et~al.}]{bai2025qwen2}
Shuai Bai, Keqin Chen, Xuejing Liu, Jialin Wang, Wenbin Ge, Sibo Song, Kai Dang, Peng Wang, Shijie Wang, Jun Tang, and 1 others. 2025.
\newblock Qwen2. 5-vl technical report.
\newblock \emph{arXiv preprint arXiv:2502.13923}.

\bibitem[{Chen et~al.(2025)Chen, Niu, Li, Liu, Zheng, Tang, Li, Xiong, and Li}]{HaluMem}
Ding Chen, Simin Niu, Kehang Li, Peng Liu, Xiangping Zheng, Bo~Tang, Xinchi Li, Feiyu Xiong, and Zhiyu Li. 2025.
\newblock Halumem: Evaluating hallucinations in memory systems of agents.
\newblock \emph{CoRR}, abs/2511.03506.

\bibitem[{Chen et~al.(2024)Chen, Liu, Huang, Wu, Liu, Jiang, Pu, Lei, Chen, Wang, Zheng, Lian, and Chen}]{survey_personalized_llm_1}
Jin Chen, Zheng Liu, Xu~Huang, Chenwang Wu, Qi~Liu, Gangwei Jiang, Yuanhao Pu, Yuxuan Lei, Xiaolong Chen, Xingmei Wang, Kai Zheng, Defu Lian, and Enhong Chen. 2024.
\newblock When large language models meet personalization: perspectives of challenges and opportunities.
\newblock \emph{World Wide Web {(WWW)}}, 27(4):42.

\bibitem[{Chhikara et~al.(2025)Chhikara, Khant, Aryan, Singh, and Yadav}]{Mem0}
Prateek Chhikara, Dev Khant, Saket Aryan, Taranjeet Singh, and Deshraj Yadav. 2025.
\newblock Mem0: Building production-ready {AI} agents with scalable long-term memory.
\newblock \emph{CoRR}, abs/2504.19413.

\bibitem[{Fang et~al.(2025)Fang, Deng, Xu, Jiang, Tang, Xu, Deng, Yao, Wang, Qiao, Chen, and Zhang}]{lightmem}
Jizhan Fang, Xinle Deng, Haoming Xu, Ziyan Jiang, Yuqi Tang, Ziwen Xu, Shumin Deng, Yunzhi Yao, Mengru Wang, Shuofei Qiao, Huajun Chen, and Ningyu Zhang. 2025.
\newblock Lightmem: Lightweight and efficient memory-augmented generation.
\newblock \emph{CoRR}, abs/2510.18866.

\bibitem[{Gutierrez et~al.(2024)Gutierrez, Shu, Gu, Yasunaga, and Su}]{HippoRAG}
Bernal~Jimenez Gutierrez, Yiheng Shu, Yu~Gu, Michihiro Yasunaga, and Yu~Su. 2024.
\newblock Hipporag: Neurobiologically inspired long-term memory for large language models.
\newblock In \emph{NeurIPS}.

\bibitem[{Hu et~al.(2025)Hu, Liu, Yue, Zhang, Liu, Zhu, Lin, Guo, Dou, Xi et~al.}]{hu2025memory}
Yuyang Hu, Shichun Liu, Yanwei Yue, Guibin Zhang, Boyang Liu, Fangyi Zhu, Jiahang Lin, Honglin Guo, Shihan Dou, Zhiheng Xi, and 1 others. 2025.
\newblock Memory in the age of ai agents.
\newblock \emph{arXiv preprint arXiv:2512.13564}.

\bibitem[{Huang et~al.(2025)Huang, Yu, Ma, Zhong, Feng, Wang, Chen, Peng, Feng, Qin, and Liu}]{llm_Hallucination}
Lei Huang, Weijiang Yu, Weitao Ma, Weihong Zhong, Zhangyin Feng, Haotian Wang, Qianglong Chen, Weihua Peng, Xiaocheng Feng, Bing Qin, and Ting Liu. 2025.
\newblock A survey on hallucination in large language models: Principles, taxonomy, challenges, and open questions.
\newblock \emph{{ACM} Trans. Inf. Syst.}, 43(2):42:1--42:55.

\bibitem[{Hurst et~al.(2024)Hurst, Lerer, Goucher, Perelman, Ramesh, Clark, Ostrow, Welihinda, Hayes, Radford et~al.}]{hurst2024gpt}
Aaron Hurst, Adam Lerer, Adam~P Goucher, Adam Perelman, Aditya Ramesh, Aidan Clark, AJ~Ostrow, Akila Welihinda, Alan Hayes, Alec Radford, and 1 others. 2024.
\newblock Gpt-4o system card.
\newblock \emph{arXiv preprint arXiv:2410.21276}.

\bibitem[{Jang et~al.(2023)Jang, Boo, and Kim}]{benchmarks_old}
Jihyoung Jang, Minseong Boo, and Hyounghun Kim. 2023.
\newblock Conversation chronicles: Towards diverse temporal and relational dynamics in multi-session conversations.
\newblock In \emph{{EMNLP}}, pages 13584--13606. Association for Computational Linguistics.

\bibitem[{Kim et~al.(2023)Kim, Kwon, Jo, and Choi}]{general_LLM}
Jiho Kim, Yeonsu Kwon, Yohan Jo, and Edward Choi. 2023.
\newblock {KG-GPT:} {A} general framework for reasoning on knowledge graphs using large language models.
\newblock In \emph{{EMNLP} (Findings)}, pages 9410--9421. Association for Computational Linguistics.

\bibitem[{Lancichinetti and Fortunato(2012)}]{lancichinetti2012consensus}
Andrea Lancichinetti and Santo Fortunato. 2012.
\newblock Consensus clustering in complex networks.
\newblock \emph{Scientific reports}, 2(1):336.

\bibitem[{Lewis et~al.(2020)Lewis, Perez, Piktus, Petroni, Karpukhin, Goyal, K{\"{u}}ttler, Lewis, Yih, Rockt{\"{a}}schel, Riedel, and Kiela}]{RAG}
Patrick Lewis, Ethan Perez, Aleksandra Piktus, Fabio Petroni, Vladimir Karpukhin, Naman Goyal, Heinrich K{\"{u}}ttler, Mike Lewis, Wen{-}tau Yih, Tim Rockt{\"{a}}schel, Sebastian Riedel, and Douwe Kiela. 2020.
\newblock Retrieval-augmented generation for knowledge-intensive {NLP} tasks.
\newblock In \emph{NeurIPS}.

\bibitem[{Li et~al.(2025)Li, Zhang, Bo, Tian, Chen, Dai, Dong, and Tang}]{li2025cam}
Rui Li, Zeyu Zhang, Xiaohe Bo, Zihang Tian, Xu~Chen, Quanyu Dai, Zhenhua Dong, and Ruiming Tang. 2025.
\newblock \href {https://openreview.net/forum?id=ACSOnSHiWe} {{CAM}: A constructivist view of agentic memory for {LLM}-based reading comprehension}.
\newblock In \emph{The Thirty-ninth Annual Conference on Neural Information Processing Systems}.

\bibitem[{Maharana et~al.(2024)Maharana, Lee, Tulyakov, Bansal, Barbieri, and Fang}]{LOCOMO}
Adyasha Maharana, Dong{-}Ho Lee, Sergey Tulyakov, Mohit Bansal, Francesco Barbieri, and Yuwei Fang. 2024.
\newblock Evaluating very long-term conversational memory of {LLM} agents.
\newblock In \emph{{ACL} {(1)}}, pages 13851--13870. Association for Computational Linguistics.

\bibitem[{Packer et~al.(2023)Packer, Fang, Patil, Lin, Wooders, and Gonzalez}]{memgpt}
Charles Packer, Vivian Fang, Shishir~G. Patil, Kevin Lin, Sarah Wooders, and Joseph~E. Gonzalez. 2023.
\newblock Memgpt: Towards llms as operating systems.
\newblock \emph{CoRR}, abs/2310.08560.

\bibitem[{Pan et~al.(2025)Pan, Wu, Jiang, Luo, Cheng, Li, Yang, Lin, Zhao, Qiu, and Gao}]{pan2025secom}
Zhuoshi Pan, Qianhui Wu, Huiqiang Jiang, Xufang Luo, Hao Cheng, Dongsheng Li, Yuqing Yang, Chin-Yew Lin, H.~Vicky Zhao, Lili Qiu, and Jianfeng Gao. 2025.
\newblock \href {https://openreview.net/forum?id=xKDZAW0He3} {Secom: On memory construction and retrieval for personalized conversational agents}.
\newblock In \emph{The Thirteenth International Conference on Learning Representations}.

\bibitem[{Pattnaik et~al.(2024)Pattnaik, George, Tripathi, Vutla, and Vepa}]{graph}
Anup Pattnaik, Cijo George, Rishabh~Kumar Tripathi, Sasanka Vutla, and Jithendra Vepa. 2024.
\newblock Improving hierarchical text clustering with llm-guided multi-view cluster representation.
\newblock In \emph{{EMNLP} (Industry Track)}, pages 719--727. Association for Computational Linguistics.

\bibitem[{Rezazadeh et~al.(2025)Rezazadeh, Li, Wei, and Bao}]{MemTree}
Alireza Rezazadeh, Zichao Li, Wei Wei, and Yujia Bao. 2025.
\newblock From isolated conversations to hierarchical schemas: Dynamic tree memory representation for llms.
\newblock In \emph{{ICLR}}. OpenReview.net.

\bibitem[{Tan et~al.(2024{\natexlab{a}})Tan, Liu, and Jiang}]{finetune-personalized-llm-2}
Zhaoxuan Tan, Zheyuan Liu, and Meng Jiang. 2024{\natexlab{a}}.
\newblock Personalized pieces: Efficient personalized large language models through collaborative efforts.
\newblock In \emph{{EMNLP}}, pages 6459--6475. Association for Computational Linguistics.

\bibitem[{Tan et~al.(2024{\natexlab{b}})Tan, Zeng, Tian, Liu, Yin, and Jiang}]{finetune-persoanlized-llm}
Zhaoxuan Tan, Qingkai Zeng, Yijun Tian, Zheyuan Liu, Bing Yin, and Meng Jiang. 2024{\natexlab{b}}.
\newblock Democratizing large language models via personalized parameter-efficient fine-tuning.
\newblock In \emph{{EMNLP}}, pages 6476--6491. Association for Computational Linguistics.

\bibitem[{Ugander and Backstrom(2013)}]{LPA}
Johan Ugander and Lars Backstrom. 2013.
\newblock Balanced label propagation for partitioning massive graphs.
\newblock In \emph{{WSDM}}, pages 507--516. {ACM}.

\bibitem[{Wang et~al.(2025{\natexlab{a}})Wang, Zhang, Zhu, Zhang, Shi, and Feng}]{retrieval}
Chengbing Wang, Yang Zhang, Fengbin Zhu, Jizhi Zhang, Tianhao Shi, and Fuli Feng. 2025{\natexlab{a}}.
\newblock Leveraging memory retrieval to enhance llm-based generative recommendation.
\newblock In \emph{{WWW} (Companion Volume)}, pages 1346--1350. {ACM}.

\bibitem[{Wang et~al.(2025{\natexlab{b}})Wang, Wu, Tan, Li, Zhong, Liu, and Zeng}]{local-global}
Zehong Wang, Junlin Wu, Zhaoxuan Tan, Bolian Li, Xianrui Zhong, Zheli Liu, and Qingkai Zeng. 2025{\natexlab{b}}.
\newblock From personal to collective: On the role of local and global memory in {LLM} personalization.
\newblock \emph{CoRR}, abs/2509.23767.

\bibitem[{Xu et~al.(2022)Xu, Szlam, and Weston}]{benchmark_msc}
Jing Xu, Arthur Szlam, and Jason Weston. 2022.
\newblock Beyond goldfish memory: Long-term open-domain conversation.
\newblock In \emph{{ACL} {(1)}}, pages 5180--5197. Association for Computational Linguistics.

\bibitem[{Xu et~al.(2025)Xu, Liang, Mei, Gao, Tan, and Zhang}]{xu2025amem}
Wujiang Xu, Zujie Liang, Kai Mei, Hang Gao, Juntao Tan, and Yongfeng Zhang. 2025.
\newblock \href {https://openreview.net/forum?id=FiM0M8gcct} {A-mem: Agentic memory for {LLM} agents}.
\newblock In \emph{The Thirty-ninth Annual Conference on Neural Information Processing Systems}.

\bibitem[{Yan et~al.(2025)Yan, Li, Qian, Lu, and Liu}]{GAM}
BY~Yan, Chaofan Li, Hongjin Qian, Shuqi Lu, and Zheng Liu. 2025.
\newblock General agentic memory via deep research.
\newblock \emph{arXiv preprint arXiv:2511.18423}.

\bibitem[{Zhang and Zhang(2025)}]{zhang2025hallucination}
Wan Zhang and Jing Zhang. 2025.
\newblock Hallucination mitigation for retrieval-augmented large language models: a review.
\newblock \emph{Mathematics}, 13(5):856.

\bibitem[{Zhang et~al.(2025{\natexlab{a}})Zhang, Yuan, and Jiang}]{associative}
Yujie Zhang, Weikang Yuan, and Zhuoren Jiang. 2025{\natexlab{a}}.
\newblock Bridging intuitive associations and deliberate recall: Empowering {LLM} personal assistant with graph-structured long-term memory.
\newblock In \emph{{ACL} (Findings)}, pages 17533--17547. Association for Computational Linguistics.

\bibitem[{Zhang et~al.(2024)Zhang, Bo, Ma, Li, Chen, Dai, Zhu, Dong, and Wen}]{zhang2024survey}
Zeyu Zhang, Xiaohe Bo, Chen Ma, Rui Li, Xu~Chen, Quanyu Dai, Jieming Zhu, Zhenhua Dong, and Ji-Rong Wen. 2024.
\newblock A survey on the memory mechanism of large language model based agents.
\newblock \emph{arXiv preprint arXiv:2404.13501}.

\bibitem[{Zhang et~al.(2025{\natexlab{b}})Zhang, Rossi, Kveton, Shao, Yang, Zamani, Dernoncourt, Barrow, Yu, Kim, Zhang, Gu, Derr, Chen, Wu, Chen, Wang, Mitra, Lipka, Ahmed, and Wang}]{survey_persoanlzied_llm_2}
Zhehao Zhang, Ryan~A. Rossi, Branislav Kveton, Yijia Shao, Diyi Yang, Hamed Zamani, Franck Dernoncourt, Joe Barrow, Tong Yu, Sungchul Kim, Ruiyi Zhang, Jiuxiang Gu, Tyler Derr, Hongjie Chen, Junda Wu, Xiang Chen, Zichao Wang, Subrata Mitra, Nedim Lipka, and 2 others. 2025{\natexlab{b}}.
\newblock Personalization of large language models: {A} survey.
\newblock \emph{Trans. Mach. Learn. Res.}, 2025.

\bibitem[{Zhao et~al.(2025)Zhao, Zhong, Sun, Hu, Liu, Li, Hu, and Zhang}]{FunnelRAG}
Xinping Zhao, Yan Zhong, Zetian Sun, Xinshuo Hu, Zhenyu Liu, Dongfang Li, Baotian Hu, and Min Zhang. 2025.
\newblock Funnelrag: {A} coarse-to-fine progressive retrieval paradigm for {RAG}.
\newblock In \emph{{NAACL} (Findings)}, pages 3029--3046. Association for Computational Linguistics.

\bibitem[{Zhong et~al.(2024)Zhong, Guo, Gao, Ye, and Wang}]{memorybank}
Wanjun Zhong, Lianghong Guo, Qiqi Gao, He~Ye, and Yanlin Wang. 2024.
\newblock Memorybank: Enhancing large language models with long-term memory.
\newblock In \emph{{AAAI}}, pages 19724--19731. {AAAI} Press.

\end{thebibliography}

\clearpage
\appendix
\section{Algorithm}
\label{app: algo}
Here we list the algorithm of Bi-Mem’s bidirectional hierarchical memory construction process in Algorithm \ref{alg:construction}, and associative retrieval mechanism in Algorithm \ref{alg:retrieval}.

\begin{algorithm}[t]
\caption{Bidirectional Hierarchical Memory Construction}
\label{alg:construction}
\begin{algorithmic}[1]
\REQUIRE Conversation history $\mathcal{C}$, similarity threshold $\tau$  
\ENSURE Calibrated hierarchical memory $\mathcal{M} = ( \mathcal{F}, \mathcal{S}', \mathcal{P})$

\STATE \textbf{// Phase 1: Inductive Process (Bottom-Up)}
\STATE $\mathcal{F} \leftarrow \text{ExtractFacts}(\mathcal{C})$ via operator $\mathcal{E}$
\STATE $G \leftarrow \text{ConstructFactGraph}(\mathcal{F}, \tau)$
\STATE $\{\bar{\mathcal{F}}_j\}_{j=1}^J \leftarrow \text{LPA}(G)$ \hfill $\triangleright$ Graph-based clustering
\STATE $\mathcal{S} \leftarrow \text{AggregateScenes}(\{\bar{\mathcal{F}}_j\})$ via operator $\mathcal{T}$
\STATE $\mathcal{P} \leftarrow \text{DistillPersona}(\mathcal{S})$ via operator $\mathcal{D}$ \hfill $\triangleright$ 5-dim persona profile

\STATE \textbf{// Phase 2: Reflective Process (Top-Down)}
\STATE $\mathcal{S}' \leftarrow \emptyset$  
\FOR{each scene $s_j \in \mathcal{S}$}
   
    \IF{$s_j$ misalign with $\mathcal{P}$}
        \STATE $\Delta s_j \leftarrow \text{Calibrate}(s_j,\mathcal{P})$ via operator $\mathcal{R}$
        \STATE $s_j' \leftarrow (s_j, \Delta s_j)$ \hfill $\triangleright$ Condition-based calibration
    \ELSE
        \STATE $s_j' \leftarrow (s_j, \emptyset)$
    \ENDIF
    \STATE $\mathcal{S}' \leftarrow \mathcal{S}' \cup \{s_j'\}$ 
\ENDFOR

\STATE \textbf{return} $\mathcal{M} = \langle \mathcal{F}, \mathcal{S}', \mathcal{P} \rangle$  
\end{algorithmic}
\end{algorithm}

\begin{algorithm}[t]
\caption{Associative Retrieval Mechanism}
\label{alg:retrieval}
\begin{algorithmic}[1]
\REQUIRE Hierarchical memory $\mathcal{M} = \langle \mathcal{F}, \mathcal{S}, \mathcal{P} \rangle$, Query $q^*$, parameters $k, m$\ENSURE Final retrieved memory set $\mathcal{M}_{ret}$\STATE \textbf{// Phase 1: Initial Hierarchical Search}\STATE Compute $a_0(x) = \text{sim}(\phi(q^*), \phi(x))$ for all $x \in \mathcal{F} \cup \mathcal{S} \cup \mathcal{P}$\STATE $\mathcal{M}_{ret} \leftarrow \text{Top-}k(\{x \mid a_0(x)\})$ \hfill $\triangleright$ Select initial seeds across layers\STATE \textbf{// Phase 2: Associative Spreading}\FOR{each element $u \in \mathcal{M}_{ret}$}\IF{$u \in \mathcal{F}$}\STATE $s \leftarrow \text{FindParent}(u)$ \hfill $\triangleright$ Bottom-up: Associate parent scene\STATE $\mathcal{M}_{ret} \leftarrow \mathcal{M}_{ret} \cup \{s\}$\ELSIF{$u \in \mathcal{S}$}\STATE $\bar{\mathcal{F}}_u \leftarrow \text{FactCluster}(u)$ \hfill $\triangleright$ Top-down: Associate constituent facts\STATE $\mathcal{F}_{top\_m} \leftarrow \text{Top-}m(\{f \in \bar{\mathcal{F}}_u \mid \text{sim}(u, f)\})$\STATE $\mathcal{M}_{ret} \leftarrow \mathcal{M}_{ret} \cup \mathcal{F}_{top\_m}$\ENDIF\ENDFOR
\STATE \textbf{return}$\mathcal{M}_{ret}$
\end{algorithmic}
\end{algorithm}

\section{Detailed Experimental Settings}
\subsection{Details for Baselines}
Here, we introduce the baseline methods in detail.
\label{app: baseline}
\begin{itemize}[leftmargin=*, itemsep=0pt, parsep=0pt, topsep=0pt, partopsep=0pt]
\item LongContext: it directly uses LLMs (no memory mechanisms) for QA tasks, integrating the historical conversations and queries into the prompt. 
\item RAG \citep{RAG}, the conversations are split into 2,048-token chunks, with the top-5 retrieved chunks used for QA tasks, following the settings of GAM \citep{GAM}.
\item A-MEM \citep{xu2025amem} is an agentic framework with structured memory, dynamic links, and semantic matching-driven evolution.
\item LightMem \citep{lightmem} is a lightweight system with a three-stage memory (sensory/short-term/long-term).
\item Mem0 \citep{Mem0} extracts memory with ADD/UPDATE/DELETE/NOOP operations and retrieves via vector similarity.
\item SeCom \citep{pan2025secom} builds denoised memory from topically coherent chunks and retrieves at the segment level.
\item CAM \citep{li2025cam} is a hierarchical memory framework that adopts an incremental overlapping clustering algorithm for construction and employs a Prune-and-Grow strategy for retrieval.
\end{itemize}

\subsection{More Implementation Details}
To implement the ablation variant ``Top-Down'' and ``Bottom-Up'' in Section \ref{sec: abla_retriever}, we set the number of selected relevant memory units to 1, 15, and 25 for the fact, scene, and persona levels, respectively, in line with mainstream configurations. Total tokens consumed to construct the hierarchical memory per user are approximately 200,000–300,000, while those for answering subsequent questions per user range from 100,000 to 20,000. The embedding model "all-MiniLM-L6-v2" is deployed on an A800, supporting both edge construction in Eq. \eqref{equ: edge} and the retrieval process in Eqs. \eqref{eq: initial_search}, \eqref{eq: associative}. Additionally, the detailed settings for our hyperparameters are listed in Table \ref{tab: hyper}. For reproduction, the data and code is available at \url{https://github.com/tohsaka-sb/Bi-Mem}.
\label{app: hyper}

\begin{table}[t!]
    \resizebox{\columnwidth}{!}{%
    \begin{tabular}{lcccc}
    \toprule
     & Single Hop & Multi Hop & Temporal&Open Domain \\
    \midrule
    $k$&35& 25 &30 &25\\
    $m$&3&3&3&3\\
    $\tau$&0.2&0.2&0.2&0.2\\
    $\alpha$&0.5&0.5&0.5&0.5\\
    $L_w$&128k&128k&128k&128k\\
    
    \bottomrule
    \end{tabular}%
    }
    \caption{Hyperparameter settings on different question types. $\alpha$ denotes the weighting factor between the dense retriever and the BM25 retriever when retrieving memories. $L_w$ represents the context window for LLM backbones.}
    \label{tab: hyper}
\end{table}

\section{More Experimental Results}
\subsection{More Sensitivity Analysis}
\label{app: sen}
To validate Bi-Mem's sensitivity to the hyperparameter $m$ (representing the number of activated facts from retrieved scenes during associative spreading activation), we present the experimental results in Table \ref{tab:sen_m}. We observe that performance peaks at $m=3$, achieving an average $F_1$ of 49.74\%. while both $m=1$ and $m=5$ lead to marginal degradation. Specifically, $m=1$ (a small value) fails to provide sufficient factual memory for complex tasks, whereas increasing $m$ to 5 may introduce redundant noise to distract the LLM from core user intent. This demonstrates that a balanced association scale can effectively retrieve optimal relevant memories for accurate question answering.
\begin{table*}[t!]
    \centering
   
    \resizebox{\textwidth}{!}{%
    \begin{tabular}{l|cc|cc|cc|cc|cc}
        \toprule
        \multirow{2}{*}{\textbf{Method}} & \multicolumn{2}{c|}{\textbf{Single Hop}} & \multicolumn{2}{c|}{\textbf{Multi-Hop}} & \multicolumn{2}{c|}{\textbf{Open Domain}} & \multicolumn{2}{c|}{\textbf{Temporal}} & \multicolumn{2}{c}{\textbf{Average}}\\
         & $F_1 \uparrow$ & $B_1 \uparrow$ & $F_1 \uparrow$ & $B_1 \uparrow$ & $F_1 \uparrow$ & $B_1 \uparrow$ & $F_1 \uparrow$ & $B_1 \uparrow$ & $F_1 \uparrow$ & $B_1 \uparrow$ \\
    \midrule
    m=1 &  52.79 & 46.95 & 34.56 & 27.83 & 25.22 & 23.01 & 51.12 & \textbf{44.77} & 47.12 & 41.25 \\
    m=5 & 53.58 & 47.82 & 37.52 & 29.98  & 29.91 & 26.03 & 53.24 & 41.03 & 49.06 & 41.73 \\
    \textbf{Bi-Mem}$(m=3)$ & \textbf{53.68} & \textbf{47.99}  & \textbf{39.17} & \textbf{31.81} & \textbf{30.99} & \textbf{26.90} & \textbf{54.44} & 41.56  & \textbf{49.74} & \textbf{42.33} \\
    \bottomrule
    \end{tabular}%
    }
    \caption{Sensitivity experiment on the hyperparameter $m$, which represents the number of extended facts in Scene\_to\_Fact when conducting associative spreading activation.}
    \label{tab:sen_m}
\end{table*}

\subsection{Case Study}
\label{app: case}
\begin{figure*}[t!]
    \centering
    \includegraphics[width=1.0\linewidth]{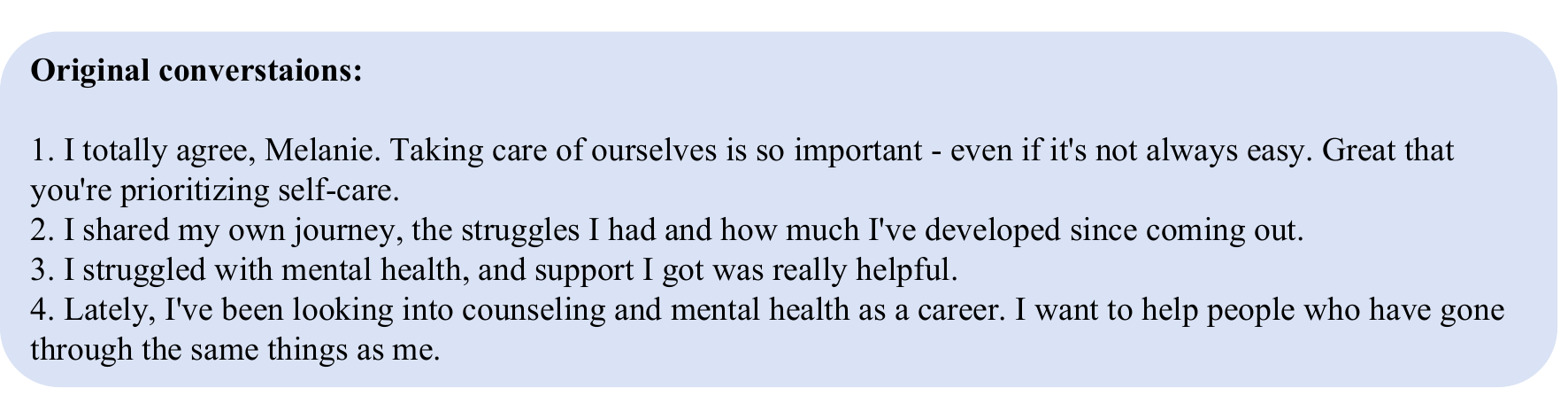}
    \caption{Case study in a cluster of original conversations.}
    \vspace{-2mm}
    \label{fig: case_converstaions}
\end{figure*}

\begin{figure*}[t!]
    \centering
    \includegraphics[width=1.0\linewidth]{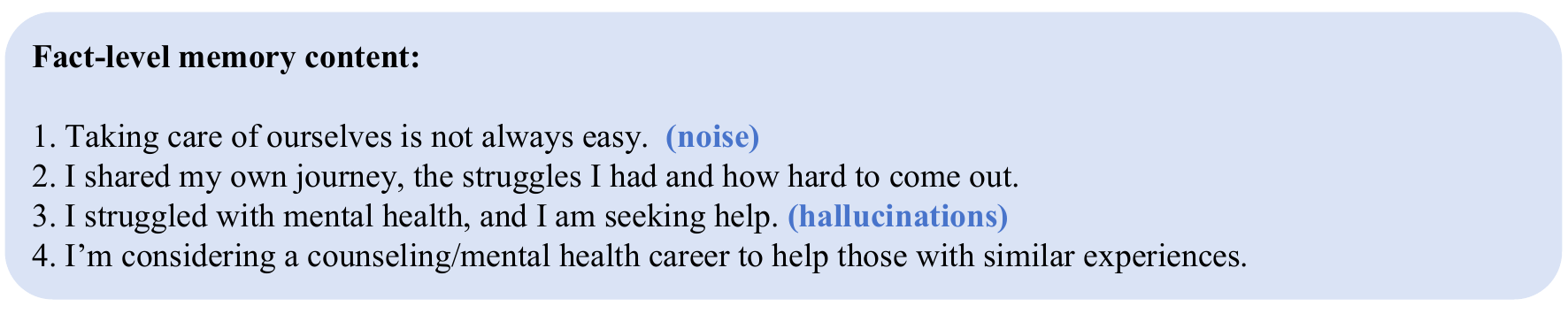}
    \caption{Case study on extracted fact-level memory content for the conversational cluster, where noise and hallucination exist.}
    \vspace{-2mm}
    \label{fig: case_facts}
\end{figure*}

\begin{figure*}[t!]
    \centering
    \includegraphics[width=1.0\linewidth]{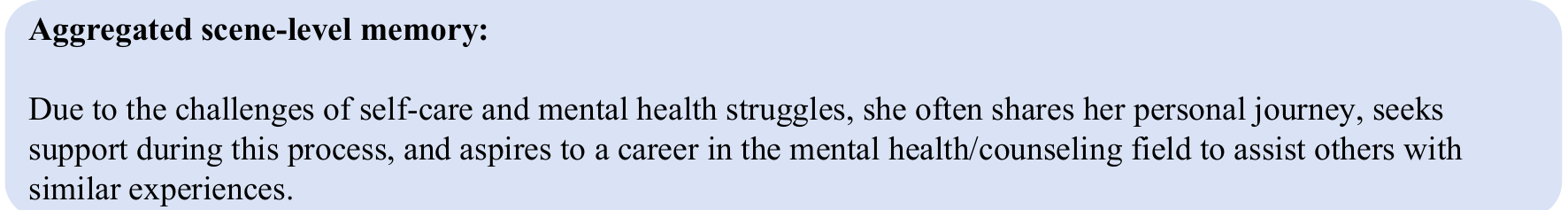}
    \caption{Case study on the aggregated scene-level memory from the fact cluster, which is misaligned with the global persona.}
    \vspace{-2mm}
    \label{fig: case_scene}
\end{figure*}

\begin{figure*}[t!]
    \centering
    \includegraphics[width=1.0\linewidth]{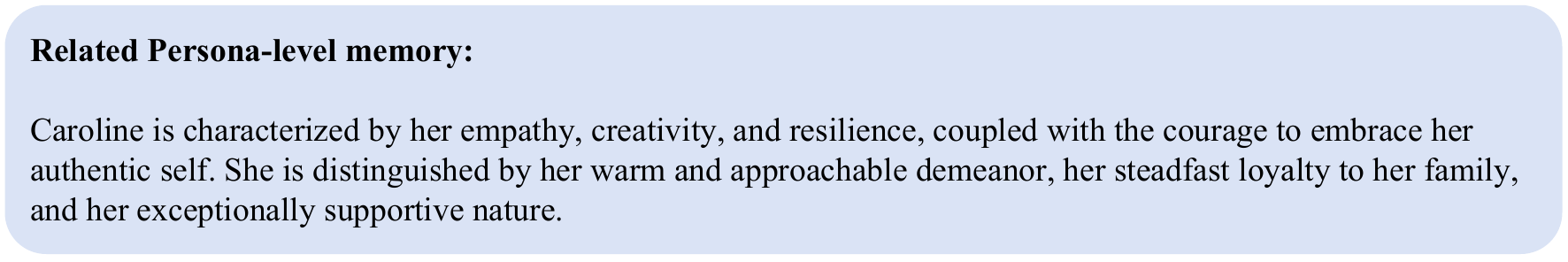}
    \caption{Case study on the global persona-level memory related to the scene.}
    \vspace{-2mm}
    \label{fig: case_persona}
\end{figure*}

\begin{figure*}[t!]
    \centering
    \includegraphics[width=1.0\linewidth]{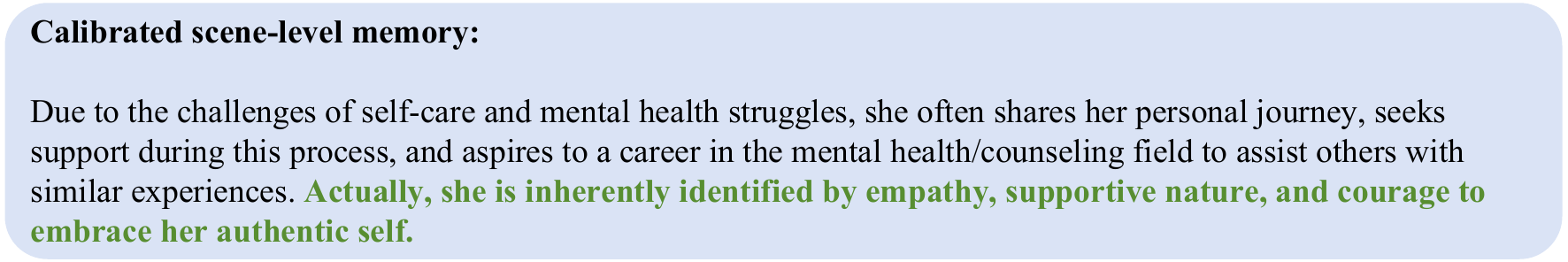}
    \caption{Case study on the calibrated scene-level memory, which is injected with the persona constraint.}
    \vspace{-2mm}
    \label{fig: case_calibarted}
\end{figure*}

To vividly illustrate the significance of bidirectional construction for hierarchical memory, we present a representative case study focusing on inductive memory construction and Reflective Calibration. As depicted in Figure \ref{fig: case_converstaions} (original conversations) and Figure \ref{fig: case_facts} (fact-level memory content), the factual information extracted from raw dialogues may contain noise (\eg ``Taking care of ourselves is not always easy'') and hallucinations (\eg ``I am seeking help''). Such inaccuracies can distort the aggregated scene-level memory (Figure \ref{fig: case_scene}), leading to misalignment with the user’s global persona-level memory—where Caroline is defined by empathy and courage to embrace her authentic self (Figure \ref{fig: case_persona}). Through Reflective Calibration, we inject the persona as a global constraint into the scene-level memory, ultimately generating a calibrated scene-level memory (Figure \ref{fig: case_calibarted}) that integrates inherent persona traits (\eg empathy and courage to embrace her authentic self) and ensures strict alignment between the scene-level memory and the core persona.

\section{Prompt Template}
Below, we present the prompts that are designed for each corresponding step of the bidirectional hierarchical memory construction.

\begin{tcolorbox}[colback=black!5!white,colframe=black!75!black,title=Fact-level memory extraction $\mathcal{E}$]

        Generate a structured fact $\text{con}_i$ for the following interaction content $c_i$ by:
        
            1. Identifying the most salient keywords (focus on nouns, verbs, and key concepts)
            
            2. Extracting core themes and contextual elements
            
            3. Creating relevant categorical tags

            Format the response as a JSON object:
            
            \{
            
                - keywords: [``keyword1'', ``keyword2'', ...],
                
                - context: ``one sentence summarizing the interaction content'',
                
                - tags: [``tag1'', ``tag2'', ...]
                
            \}

\end{tcolorbox}

\begin{tcolorbox}[colback=black!5!white,colframe=black!75!black,title=Scene-level memory aggregation $\mathcal{A}$]
     
        You are a scene synthesizer specialized in factual comprehension..
        
        Task: Summarize a cluster of related factual memories into a coherent 'Scene Memory'.
        
        Input factual memories:
        {facts\_content}
        
        Instructions:
        
        1. Identify the core theme connecting these facts.
        
        2. Generate a descriptive summary capturing the progression of conversational facts.
        
        3. Extract key entities and topics.
        
        Format the response as a JSON object:
        
        \{
        
            - scene\_memory: ``A comprehensive summarized scene'',
            
            - keywords: [``keyword1'', ``keyword2'', ...],
            
            - tags: [``tag1'', ``tag2'', ...]
            
        \}

\end{tcolorbox}

\begin{tcolorbox}[colback=black!5!white,colframe=black!75!black,title=Persona-Level Memory Distillation $\mathcal{D}$]
       
        You are a persona synthesizer specialized in psychological and behavioral analysis.
        
        Task: Create a COMPREHENSIVE User persona based on the provided scene memories.
        
        Input Scenes:
        {all\_scenes\_content}
        
        Instructions:
        
        1. Analyze these scenes deeply. Look for patterns in behavior, emotion, and choices.
        
        2. For each dimension below, write a DETAILED paragraph (5-10 sentences). Do not be brief.
        
        3. Use specific examples from the scenes to support your analysis.
        
        Format the response as a JSON object:
        
        \{
        
            - basic\_info: ``Detailed background...'',
            
            - interests: ``Comprehensive list of hobbies and how they engage with them...'',
            
            - personality: "In-depth personality analysis...'',
            
            - values: ``Core beliefs and motivations...'',
            
            - relationships: ``Detailed social dynamics...''
            
        \}
   
\end{tcolorbox}

\begin{tcolorbox}[colback=black!5!white,colframe=black!75!black,title=Reflective Calibration.]

        You are a scene memory calibrator. Your goal is to align the given scene to the user's persona.
        
        Persona-level memory:
        {user\_persona}
        
        Scene-level memory:
        {current\_scene}
        
        Instructions:
        
        1. Read the persona-level memory to understand the user's key interests, values, and traits.
        
        2. Check the current scene-level memory. Does it fail to mention any specific connection to the user persona that is likely present in the scene?
        
        3. If yes, add a compensatory condition to append to the original scene. This addition should explicitly align the scene to the persona (e.g., ``This aligns with her interest in ...'').
        
        4. CRITICAL: DO NOT REWRITE the existing summary. ONLY generate text to ADD.
        
        5. If the current summary is already perfect, return an empty string for ``added condition''.
        
        Format the response as a JSON object:
        
        \{
        
            - needs\_calibration: true/false,
            
            - added condition: ``Text to add (or empty string) as a condition'',
            
            - reason: The reason why you decide to calibrate.''
            
        \}

\end{tcolorbox}

\clearpage


\end{document}